\documentstyle[twoside,fleqn,espcrc2,epsf]{article}

\title{Free Meson Spectral Functions on the Lattice
\thanks{This work is supported by 
BMBF under grant No.06BI102 and DFG under grant FOR 339/2-1. P.P. is 
Goldhaber fellow supported by the contract DE-AC02-98CH10886 with the U.S. Department of Energy.
}}

\author{S. Stickan\rlap,
\address{Fakult\"{a}t f\"{u}r Physik, Universit\"{a}t Bielefeld, D-33615 Bielefeld, Germany}
F. Karsch\rlap,$\;^{\rm a}$
E. Laermann\rlap,$\;^{\rm a}$
P. Petreczky\address{Nuclear Theory Group, Physics Department Brookhaven National Laboratory, Upton NY 11973}
}

\begin{document}

\begin{abstract}
We present results from an analytic calculation of thermal meson spectral 
functions in the infinite temperature (free field) limit. We compare 
spectral functions for various lattice fermion formulations 
used at present in studies
of in-medium properties of hadrons based on the maximum entropy method (MEM).
In particular, we will present a new calculation of spectral functions
performed with extended quark sources.
\vspace{1pc}
\end{abstract}

\maketitle
\vspace*{-10cm}\hspace*{12.2cm} \begin{minipage}{5cm}{\Large BI-TP 2003/28\\BNL-NT-03/29}\end{minipage} \vspace*{7cm}

\section{Introduction}
\label{sec:intro}

The modification of in-medium properties of mesons is one of the crucial
concepts explored in experimental studies of hot and dense matter in
created in heavy ion collisions. It is the hope that the creation of
a dense matter and eventually also the quark gluon plasma can be 
unambiguously demonstrated through the study of modifications
generated in the spectrum of heavy ($J/\psi$-suppression) as well as
light quark bound states (shift of the $\rho$-meson mass, resonance
broadening, thermal dilepton enhancement,..).

Lattice calculations can contribute to the theoretical discussion of these
burning experimental questions through the calculation of thermal 
spectral functions, which control the structure of hadron correlation
functions at finite temperature. The reconstruction of spectral functions
from hadron correlation functions calculated on the lattice at finite 
temperature is based on a statistical tool, the Maximum Entropy Method 
(MEM) \cite{c1,c7,c8,c9} 
which has been explored and tested over recent years. We will discuss 
here results of an analytic calculation, which becomes feasible in the
infinite temperature limit which corresponds to a free fermion theory.
In addition to calculations on isotropic and anisotropic lattices based on 
hadronic currents constructed from point sources 
we will present a new calculation of spectral functions
performed for extended quark sources. 

\section{Free Lattice Spectral Functions}
\label{sec:SPFf}

\begin{figure}[t]
\centerline{\epsfxsize=6.8cm\epsfbox{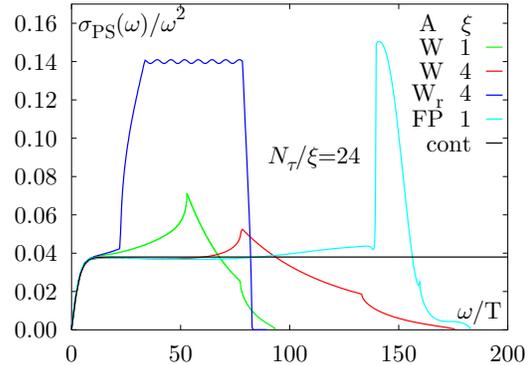}}
\vspace{-5mm}
\caption{The free pseudo-scalar SPF for the Wilson action (W) on 
isotropic ($\xi=1$) and anisotropic ($\xi=4$) lattices, the Wilson action 
with the choice $r=1/\xi=4$ ($W_r$) and a truncated fixed-point action (FP). 
A part of the SPF for $W_r$ has been cut out at intermediate energies for 
better visibility.}
\label{fig:PSPF}
\end{figure}

The calculation of hadronic spectral functions (SPF) on the lattice starts from 
the Euclidian correlation function of hadronic currents, $G_H(\tau)$, with
given quantum numbers, $H$. In the infinite temperature limit these correlation
functions (CF) will approach the free fermion CF, which can be expressed in 
terms of the free lattice quark propagator $S_L$. We will discuss here 
analytic calculations for spectral functions determined from correlation
functions defined with currents constructed from point-like (p)
as well as extended (exponentially smeared $e\equiv \exp(-x^2)$ or 
fuzzed $R=1$) sources \cite{c2,c6}. Furthermore, we will present results
on isotropic as well as anisotropic lattices. The anisotropy in
the spatial ($a$) and temporal ($a_\tau$) lattice spacings is controlled by 
the anisotropy parameter $\xi\equiv a/a_\tau$. While the calculations
on anisotropic lattices have been performed for the case of a Wilson
action with $r$-parameter being set to unity as well as for the case of
Wilson fermions where the time-like 
$r$-parameter is linked to the anisotropy $r=1/\xi$. This latter choice
is often used in heavy quark simulations and has also been used in the MEM
analysis of heavy quark bound states \cite{c2}.
In addition we present results from calculations with
a truncated fixed-point action \cite{c5}. In momentum space the Euclidean time
correlation functions for free fermions is then given by,

\begin{eqnarray}
  G^L_H(\tau)\hspace*{-10pt}&=&\hspace*{-12pt}\int\hspace*{-4pt}w_{s}({\bf k}) {\rm Tr}\hspace*{-2pt}\left[ {
      \Gamma_HS_L(\tau,{\bf k})\Gamma_H^\dagger S_L^\dagger(\tau,{\bf k})
}\right]{\rm d}^3k\quad\nonumber\\
  &=&\hspace*{-10pt}\int  \;\sigma_H^L(\omega,T) \frac{{\rm ch}(\omega(\tau-1/2T))}{{\rm sh}(\omega/2T)} \;{\rm d}\omega \;,\label{eq:frecf2}
\end{eqnarray}
\begin{eqnarray}
\label{eq:extmes}
w_s({\bf k})=
\left\{{ \begin{array}[h]{ l } 1 \\ \exp({\bf k}^2/4) \\ \exp({\bf k}^2/2) \\ \cos({\bf kR}) \end{array} }\right.\;\;\mbox{for }s=
\left\{{ \begin{array}[h]{ l } pp\\ ep \\ ee \\ R=1\end{array} }\right.,
\end{eqnarray}
where $\Gamma_H$ is a product of $\gamma$-matrices that fixes the 
quantum numbers $H$.
The crucial step in Eq.(\ref{eq:frecf2}) is the second equality where
the 3-dimensional momentum integral has been replaced by a 1-dimensional
integral over the energy $\omega$. Two further integrations are hidden
in the definition of the spectral function $\sigma^L_H(\omega, T)$. 
We note that
a separation of the dependence on Euclidean time $\tau$ in the  integration 
kernel, 
\begin{equation}
K(\omega,\tau)={\rm ch}(\omega(\tau-1/2T))/{\rm sh}(\omega/2T) \; ,
\end{equation}
which kept its continuum form, and the lattice spectral function 
$\sigma^L_H(\omega, T)$, which is $\tau$-independent and carries the entire
cut-off dependence of the CFs, could be achieved. 
This makes the introduction of a lattice kernel, which
has been used in some numerical studies based on the MEM analysis \cite{c1,c7} 
questionable. The
analytic calculations are presented in \cite{c4} for point-like sources
and the truncated fixed-point action. The resulting spectral functions
obtained in these cases on lattices with different anisotropies are
shown in Fig.~\ref{fig:PSPF}. 

A closer look at the resulting SPFs shows that they all agree quite
well with the continuum result,
\begin{equation}
\sigma_{\rm cont} (\omega, T) = \frac{3}{8\pi^2}\;\omega^2\;\tanh\left( {\omega/T }\right)
\end{equation}
in the IR regime. This agreement extends
to larger energies for the Wilson action on anisotropic lattices and even
more so for the truncated fixed-point action \cite{c5}. 
The choice $r=1/\xi$ apparently leads to a very abrupt deviation from this
continuum like behavior and leads to large cut-off effects for 
$\omega /T \ge 10$. This makes its use for the heavy quark spectroscopy 
at high temperature questionable.
The other actions also show the influence of lattice artefacts in the UV 
region. Lattice artefacts show up as peaks and cusps which
can directly be related to the sudden restriction of the available momentum 
phase space near the corners of the Brillouin zone for the quark dispersion relation. 
A close relation between the quark dispersion relation and the structure
of SPFs thus is obvious in the free theory.

\begin{figure}[t]
\centerline{\epsfxsize=6.8cm\epsfbox{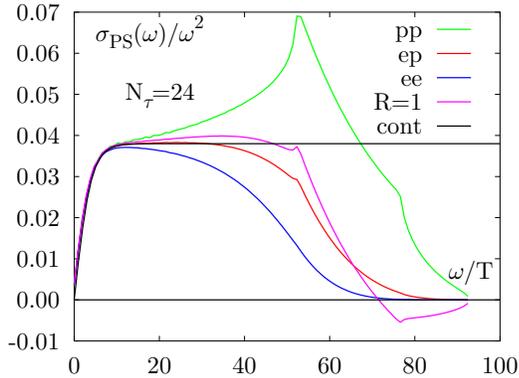}}
\vspace{-5mm}
\caption{The pseudo-scalar SPF calculated on isotropic lattices 
using for different extended meson operators 
as defined in Eq.(\ref{eq:extmes}).}
\label{fig:extmes}
\end{figure}

\section{Extended Sources}

To enhance the overlap with the ground state excitation in a given quantum 
number channel, extended operators are often used. Above $T_c$, the 
existence of a well separated ground state is, however, not guaranteed and 
it thus is questionable whether extended operators can be of any use at all.
Nevertheless, extended operators have been used in the MEM analysis of
meson correlation functions\cite{c2}. As they reduce to point like operators in the
continuum limit, when their extend is kept fixed in lattice units, they may
indeed also be suitable for studies of spectral functions. We thus have
analyzed some of them also in the free field limit. 
The SPFs of extended operators can be easily obtained using the so 
called ``binning procedure'' (see \cite{c4} for details). Some results are 
shown in Fig.~\ref{fig:extmes}. As can be seen the UV regime indeed is 
strongly suppressed and a good agreement with the continuum spectral
function can be achieved of a wide range of energies up to the region 
where the point-like currents lead to large contributions from the
``Wilson doublers''. This suppression is more effective for operators with larger extend, as can be observed with the ($ee$) smeared operator which in the considered limit can also be viewed as a broader smeared  exponential ($ep$) operator.
We note, however, that the use of extended sources
requires tuning of parameters and 
care has to be taken that the low energy part does not get to be
suppressed too, which happen easily with the ($ee$) operator. Some versions of extended operators also violate the
positivity of the spectral functions, which enters as a basic assumption in 
the MEM analysis \cite{c1}. The fuzzed ($R=1$)-source has this problem for $\omega/T\ge70$. 
Moreover, the exponentially smeared sources are not gauge invariant and gauge fixing becomes necessary. For the application in simulations this additional computational overhead has to be compared with the improvement which could be achieved by using improved fermion actions or by increasing $N_\tau$.

\section{Conclusions}

The analysis of the free lattice SPF provides basic information on the
structure of spectral functions reconstructed in the interacting case
from a statistical analysis based on {MEM}. It also can provide guidance
for a suitable choice of the default model in the MEM analysis and the
choice of an appropriate integration kernel in the spectral representation.
In the free field limit
lattice artefacts only lead to a modification of the SPF, the integration
kernel keeps its continuum form.
Extended meson operators lead to an efficient suppression of the UV part of 
the SPFs. This, however,  requires a careful choice of parameters in order not 
to influence also the IR part of the spectral functions which at finite 
temperature is not well separated from the short distance part of 
Euclidean correlation functions.

\end{document}